# Effect of Li doping on magnetic and transport properties of $CoV_2O_4$ and $FeV_2O_4$

Prashant Shahi[1], R. Singh[1],Shiv Kumar[2], D. K. Dubey[1,3], D. N. Singh[1,3], A.Tiwari[4], A.Tripathi[4], A. K. Ghosh[2] and Sandip Chatterjee[1,*]

[1]Department of Physics, Indian Institute of Technology (Banaras Hindu University), Varanasi-221 005, India

[2]Department of Physics, Banaras Hindu University, Varanasi-221 005, India

[3]Department of Physics, National institute of Technology Durgapur, India

[4]Department of Physical Sciences, School of Chemical and Physical Sciences, SikkimUniversity, Sikkim, India

## Abstract

The structural, magnetic and transport properties have been studied of Li doped $CoV_2O_4$ and $FeV_2O_4$. Li doping increases the ferri-magnetic ordering temperature of both the samples but decreases the spin-glass transition temperature of $CoV_2O_4$. The Li-doping decreases the V-V distance which in effect increases the A-V coupling. Thus the increased A-V coupling dominate over the decrease in A-V coupling due to doping of non-magnetic Li.

*__Corresponding author__ *e-mail id*: schatterji.app@iitbhu.ac.in
*Ph.No. & Fax No.: +91-5426701913*

# Introduction

Transition- metal- oxide (TMO) spinels is becoming the centre of attraction with significant quality of complex interaction among charge, spin and orbital degrees of freedom which gives it some interesting properties. Importantly, vanadium spinel oxides $AV_2O_4$ (A $=Fe^{2+}$, $Mn^{2+}$, $Co^{2+}$, $Zn^{2+}$, $Mg^{2+}$), where $A^{2+}$ and $V^{3+}$ ions occupy the tetrahedral (A site) and octahedral (B site) sites, respectively, have two 3d electrons in the triply degenerate $t_{2g}$ states at $V^{3+}$ site. This whole arrangement is becoming the centre of attention and also their fascinating magnetic and orbital order [1–7]. When A site is replaced by some non magnetic ion [Li, Zn, Mg, Cd] it shows many interesting properties. Likewise $ZnV_2O_4$ goes from cubic to tetragonal state at T=50 K and orders antiferromagnetically at T= 42 K [8] and $LiV_2O_4$ shows metallic behavior and it is a first transition metal oxide which shows heavy fermion behavior and remains in cubic phase in its whole temperature range [9-11]. On the other hand, when A site is replaced by Magnetic ions, different properties emerges. In this way the $MnV_2O_4$ exempts magnetic phase transition from the collinear ferrimagnetic order to the non-collinear canted one at 53 K with associations to the anti-ferro-orbital order of vanadium $t_{2g}$ orbitals [3, 12]. Whereas, $CoV_2O_4$ exhibits two magnetic phase transitions at 142 K and 59 K without any structural phase transitions [13, 14].

Moreover, $FeV_2O_4$ is a unique compound among such spinel vandium oxides comprising both $Fe^{2+}$ and $V^{3+}$ ions with orbital degrees of freedom; $Fe^{2+}$ ion at the tetrahedral site (A-site) having three 3d electrons in the doubly degenerate $e_g$ states. Very recent development shows structural and magnetic properties of the spinel $FeV_2O_4$ [11] exhibiting successive structural transitions from cubic to compressed tetragonal with the lattice constants of $c < a$ at $T_{s1} \sim 140$ K which is due to the cooperative Jahn–Teller effect of $FeO_4$ and from tetragonal to orthorhombic transition accompanied by a ferrimagnetic transition at $T_{s2} \sim 110$ K, and from orthorhombic to elongated tetragonal with $c > a$ at $T_{s3} \sim 60$ K with decreasing temperature for polycrystalline samples. It has also been reported that the Jahn–Teller effects and the relativistic spin–orbital coupling play an important role in the determination of the orbital states of $Fe^{2+}$ and $V^{3+}$ ions at low temperatures which were concluded with the help of single crystal x ray diffraction experiments [6]. Recent reports of NMR and neutron diffraction of $FeV_2O_4$ indicate that its structure is changing to non-collinear ferrimagnetic state at 60 K from collinear state, where the $V^{3+}$ moments become mounted along the {111} directions [7, 15]. This latter transition is marked by a step in magnetization, a peak in heat capacity, an anomaly in the dielectric constant, and the appearance of polarization. It was found that the application of a magnetic field shifts all these signatures associated to $T_{s3}$ to higher temperatures, while it also clearly affects the value of the polarization, revealing a significant magneto electric coupling. It is suggested that the presence of canted spins in the triangular structure below $T_{s3}$ could be responsible for the appearance of ferroelectricity [16].

The $AV_2O_4$ system also approaches the itinerant-electron limit with decreasing V-V separation [17.18]. The predicted critical separation for metallic behavior is $R_c$ = 2.94 $A^0$[19]. A Recent study on $FeV_2O_4$ and $CoV_2O_4$ shows that with increasing pressure the V-V separation decreases and due to which there is a delocalization of charge carriers in $FeV_2O_4$ and it induces metallic behavior in $CoV_2O_4$ [14]. The same effect is also shown by chemical pressure by doping Co at the site of $MnV_2O_4$ [20]. Recently it is shown that in $FeV_2O4$, $CoV_2O_4$ and $MnV_2O_4$ the magnetic transition temperature suppressed and activation energy decreases as $Zn^{2+}$ (non magnetic ion) increases at the A site [21-23]. Furthermore, $Li^{1+}$ is also a non magnetic ion and its size is comparable to Mn, Zn, Fe and Co but $Li^{1+}$ has no 3d electron in their outer shell unlike Fe and Co. Therefore, it will be interesting to investigate the magnetic and transport properties by doping Li in $CoV_2O_4$ and $FeV_2O_4$ at the A site.

Furthermore, soft magnetic materials are center to nearly every aspect of modern electrical and electronics technology because of their ability to concentrate and to shape magnetic flux with great efficiency. The most important Characteristics desired for essentially all soft magnetic applications are high saturation induction, high permeability, low coercivity, and low core loss .So these Materials are also good subject for these application.

In this paper, we have investigated magnetic and transport properties of $Li_xCo_{1-x}V_2O_4$ ($0 \leq x \leq 0.2$) and $Li_xFe_{1-x}V_2O_4$($0 \leq x \leq 0.1$) and found that with increasing the Li content at Co and Fe site the Ferrimagnetic transition temperature increases for bothe the systems, the spin glass magnetic transition is suppressed in $CoV_2O_4$ and the systems are moving towards iterant electron limit due to decrease in the V-V distance.

## Experiment

The polycrystalline $Li_xCo_{1-x}V_2O_4$ (0≤x≤0.2) and $Li_xFe_{1-x}V_2O_4$(0≤x≤0.1) samples used in this study were prepared by solid state reaction method. Appropriate ratio of $Li_3VO_4$, Fe, $Fe_2O_3$, CoO $V_2O_3$ and $V_2O_5$ were grounded thoroughly and pressed into pellets. The pellets were sealed in evacuated quartz tube and heated at 1050°C for 40 hours for $Li_xFe_{1-x}V_2O_4$ (0≤x≤0.1) and at 900°C for $Li_xCo_{1-x}V_2O_4$ (0≤x≤0.2). The X-ray powder diffraction experiment has been performed using Rigaku Mini Flex II DEXTOP X-ray Diffractometer with Cu-kα radiation. Magnetic measurement was done using MPMS SQUID (Quantum Design) magnetometer with the bulk samples. Ac-susceptibility measurement were done using lock in amplifier SRS830 by homemade setup and standardized with YBCO superconducting sample. Fourier Transform Infrared Spectroscopy measurements have been done using Spectrum 65 FTIR spectrometer (Perkin Elmer Instruments, USA) in the range of 500 to 4000 $cm^{-1}$. Resistivity measurements have been done via four probe method.

## Result and Discussion

Figure 1 shows the X-ray diffraction (XRD) pattern for different Li doped samples. All peaks are indexed as Fd-3m space group indicating our samples are of pure phase. Inset of the figure 1 shows variation of lattice parameter with Li Doping, obtained from the Rietveld refinement of the XRD data. It is observed that with increasing the Li concentration at the Fe and Co site the lattice parameters decrease linearly following the Vegard's law. The ionic size of Li is 0.73Å which is comparable to Co (0.72Å) and smaller than Fe (0.77Å). But in both the cases lattice parameters are decreasing which might be due to the fact that Co have seven 3d electrons and Fe have six 3d electrons due to which there is strong coulomb repulsion between Co/Fe and oxygen 2p electrons but due to lithium doping the coulomb repulsion decreases as it has no 2p and 3d electrons in its outer shell. The parameters obtained from Reitveld refinement are given in the Table 1. Figure 2 and figure 3 show the variation of magnetization with temperature for $Li_xCo_{1-x}V_2O_4$ (0≤x≤0.2) and $Li_xFe_{1-x}V_2O_4$ (0≤x≤1) which shows that as we are increasing the Li content at Co and Fe site the ferrimagnetic transition temperature also increasing but the second magnetic transition is suppressed by Li doping in $CoV_2O_4$. We have also measured ac-susceptibility at 100 Hz for $Li_xCo_{1-x}V_2O_4$ (0≤x≤2) and $Li_xFe_{1-x}V_2O_4$ (0≤x≤1) which is shown in figure 4 and figure 5 respectively. Similar behavior is observed for both the cases that with increasing Li content at A site the ferrimagnetic transition temperatures increase. This might be due to the fact that shrinkage of the lattice parameter with Li doping, increases the exchange interaction between the $A^{2+}$ and $V^{3+}$ through oxygen which enhances the ferrimagnetic ordering temperature. On the other hand, the saturated moment, which is estimated from the isothermal magnetization curve (shown in Figure 6), decreases monotonously upon Li doping. Since Li ions are non-magnetic, it is known that V spins tend to align anti-parallel to each other when the coupling between A and B sub-lattices is absent. Therefore the moment of the V sub-lattice also decreases with increasing Li content. As a result it decreases monotonously. It is important to compare these with the earlier studies on $Zn_xMn_{1-x}V_2O_4$ [23] and $Co_{1-x}Zn_xV_2O_4$ [22] because Zn(0.74Å) is also a non magnetic ion. Moreover, the sizes of Li (0.73Å), Co (0.72Å) and Zn (0.74Å) are comparable. In case of $Zn_xMn_{1-x}V_2O_4$ it is found that as the Zn content increases the magnetic transition temperature decreases which is different than that in Li doped samples. With Zn doping two cases arises, the lattice parameters decrease with increasing Zn content due to which the V-V distance decreases so that coupling between A and V sites increase which tries to increase the magnetic transition temperature. But doping of non magnetic ion at A site decreases the A-V coupling due to which V-V spins try to align anti-parallel. As a result, the magnetic transition is suppressed [23]. The same case also arises in case of $Co_{1-x}Zn_xV_2O_4$ [22]. The second effect is more dominant in case of Zn doping but in our case the first effect is dominating due to which the magnetic transition temperature increases with Li doping. Similar behavior is observed in $Mn_{1-x}Co_xV_2O_4$ system where with increasing magnetic Co content the ferrimagnetic transition temperature increases along with the decrease of V-V distances [20]. In the present systems it might be the case that non-magnetic Li doping converts some $V^{3+}$ into $V^{4+}$ which might be the reason of increasing the A-V coupling. From Figure 2 we have found that the spin glass behavior, which is present in case of $CoV_2O_4$ at 59 K is also suppressed with $Li^{1+}$ doping. This may be due to the weakening of magnetism at the $Co^{2+}$ sub-lattice due to doping of non magnetic $Li^{1+}$ which may weaken the coupling between $Co^{2+}$ and $V^{3+}$ sub-lattice and due to which the spin glass behavior is suppressed with Li doping. Opposite behavior is observed when non-magnetic Zn is doped in $Cd_{1-x}Zn_xV_2O_4$ system [24]. Zn doping increases the V-V distance which in effect induces the site disorder. Figure 7 shows the variation of resistivity with temperature for all samples. As $FeV_2O_4$ belongs to a Mott Insulator regime and $CoV_2O_4$ is lying at the itinerant electron limit therefore from inset of the figure 7 it can be mentioned that with increasing Li content at the A site the system moves towards itinerant electron side along with the decrease of V-V distance. To see the chemical pressure effect on both the systems we have plotted all our result as a function of $1/R_{V\text{-}V}$. Figure 8 shows the absorbance spectrum of $Li_xCo_{1-x}V_2O_4$ and $Li_xFe_{1-x}V_2O_4$ in the mid-infrared region. The peak observed below ~ 1000 $cm^{-1}$ corresponds to one of the four phonon modes expected for cubic spinals. The charge gap is estimated from the intersection of the linear extrapolation of the absorption edge and the frequency axis. We have also calculated the activation energy from ln(ρ) vs 1000/T plot and plotted the charge gap and activation energy with respect to $R_{V\text{-}V}$ in figure 9. Both variations show the same nonmonotonic dependence of the gap as a function of chemical pressure as found in case of $ZnV_2O_4$ with both external and chemical pressure [25].

## Conclusion

We have found that as we are increasing the Li content at A site the V-V distance decreases and due to that ferimagnetic transition temperature increases and the whole system is moving towards iterant electron behavior. So by tuning the V-V distance either by external pressure or chemical pressure we can tune the magnetic and transport properties of $AV_2O_4$ samples which is very important for application point of view. Both the activation energy and charge gap show the non-monotonic dependence on chemical pressure.

**Acknowledgement**

SC is grateful to DST (Grant No.: SR/S2/CMP-26/2008), CSIR (Grant No.: 03(1142)/09/EMR-II) and BRNS, DAE ((Grant No.: 2013/37P/43/BRNS) for providing financial support. PS is grateful to CSIR for providing financial support.

Table 1. Structural parameters(lattice parameters, bond engths) of $Fe_{1-x}Li_xV_2O_4$(0<x<0.1) and $Co_{1-x}Li_xV_2O_4$(0<x<0.2) samples obtained from Rietveld refinement of X-ray diffraction data

| Sample Name | a(Å) | d(V-V)($A^0$) |
|---|---|---|
| $CoV_2O_4$ | 8.4075 | 2.9725 |
| $Co_{0.95}Li_{0.05}V_2O_4$ | 8.3993 | 2.9696 |
| $Co_{0.9}Li_{0.1}V_2O_4$ | 8.3908 | 2.9666 |
| $Co_{0.8}Li_{0.2}V_2O_4$ | 8.3746 | 2.9609 |
| $FeV_2O_4$ | 8.4517 | 2.9881 |
| $Fe_{0.95}Li_{0.05}V_2O_4$ | 8.4429 | 2.9850 |
| $Fe_{0.9}Li_{0.1}V_2O_4$ | 8.4341 | 2.9819 |

## Figure Captions:

1. X-ray diffraction pattern with Reitveld refinement for Li doped $CoV_2O_4$ and $FeV_2O_4$ samples at 300K. The inset shows the variation of lattice parameters with Li concentration

2. Temperature variation of magnetization for $Co_{1-x}Li_xV_2O_4$ [where x=0, 0.05, 0.1 and 0.2 spinels.at H=5000 Oe].

3. Temperature variation of magnetization for $Fe_{1-x}Li_xV_2O_4$ [where x=0, 0.05, and 0.1 spinels .at H=5000 Oe] . Inset shows the plot of dM/d T vs.T indicating transitions.

4. Temperature dependence of AC magnetization measured in fields with 100 Hz frequency for $Co_{1-x}Li_xV_2O_4$ [where x=0, 0.05, 0.1 and 0.2 ] around $T_C$. Inset shows the Variation of $T_C$ with respect to $1/R_{V-V}$.

5. Temperature dependence of AC magnetization measured in fields with 100 Hz frequency for $Fe_{1-x}Li_xV_2O_4$ [where x=0, 0.05, and 0.1] around $T_C$. Inset shows the Variation of $T_C$ with respect to $1/R_{V-V}$.

6. The isothermal field dependence of the magnetization at 2 K for $Fe_{1-x}Li_xV_2O_4$ [where x=0, 0.05, and 0.1] and $Co_{1-x}Li_xV_2O_4$ [where x=0, 0.05, 0.1 and 0.2 ], and at 80 K for $Co_{1x}Li_xV_2O_4$[where x=0, 0.05, 0.1 and 0.2 ] respectively.

7. The temperature dependences of resistivity for $Fe_{1-x}Li_xV_2O_4$ and $Co_{1-x}Li_xV_2O_4$.

8. Room temperature Absorbance Spectra of $Co_{1-x}Li_xV_2O_4$ [where x=0, 0.05, 0.1 and 0.2 ] and $Fe_{1-x}Li_xV_2O_4$ [where x=0, 0.05 and 0.1].

9. Charge gap and activation energy variation as a function of inverse V-V distance for $Fe_{1-x}Li_xV_2O_4$[where x=0, 0.05 and 0.1].and $Co_{1-x}Li_xV_2O_4$ [where x=0, 0.05, 0.1 and 0.2 ].

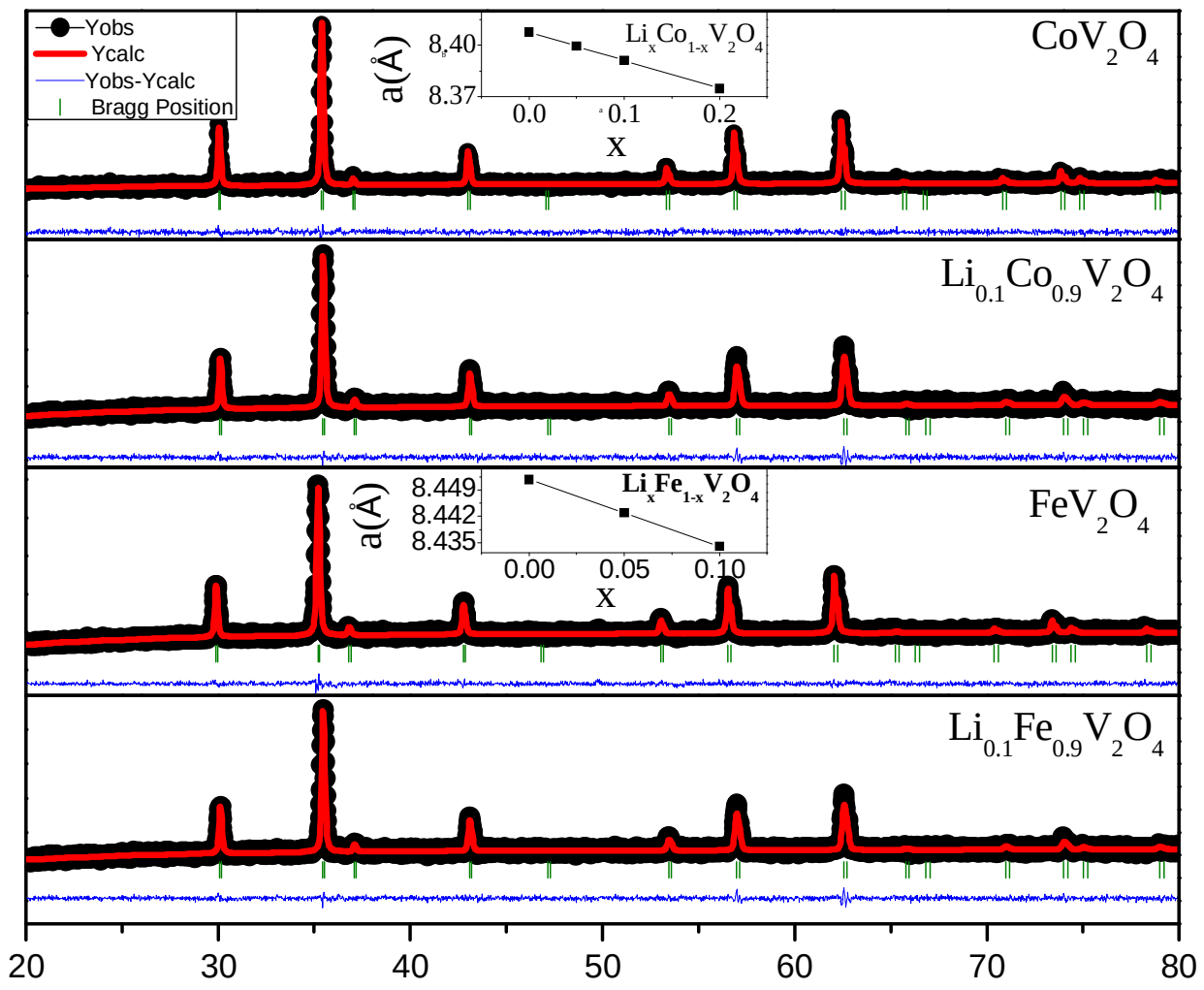

Figure 1

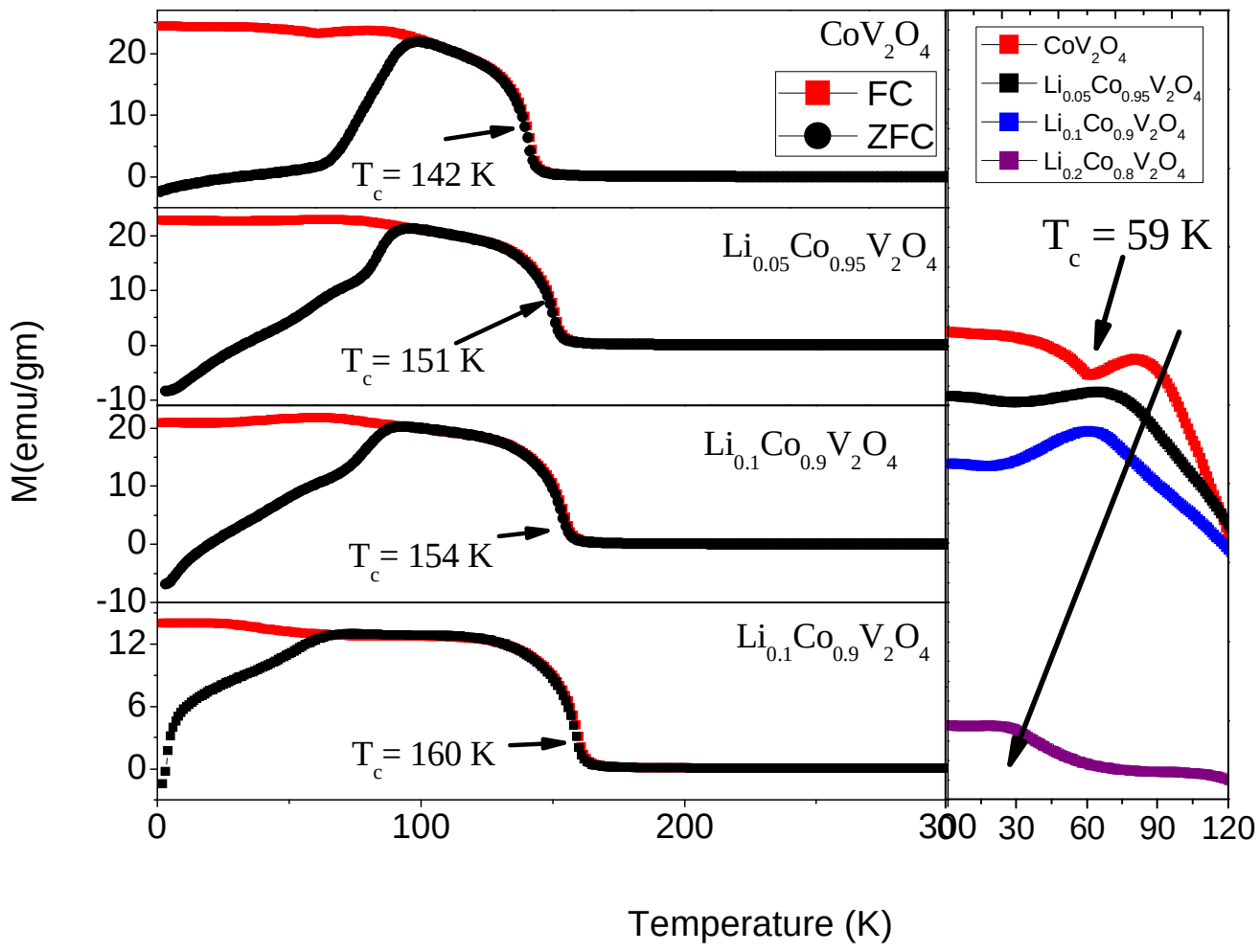

Figure 2

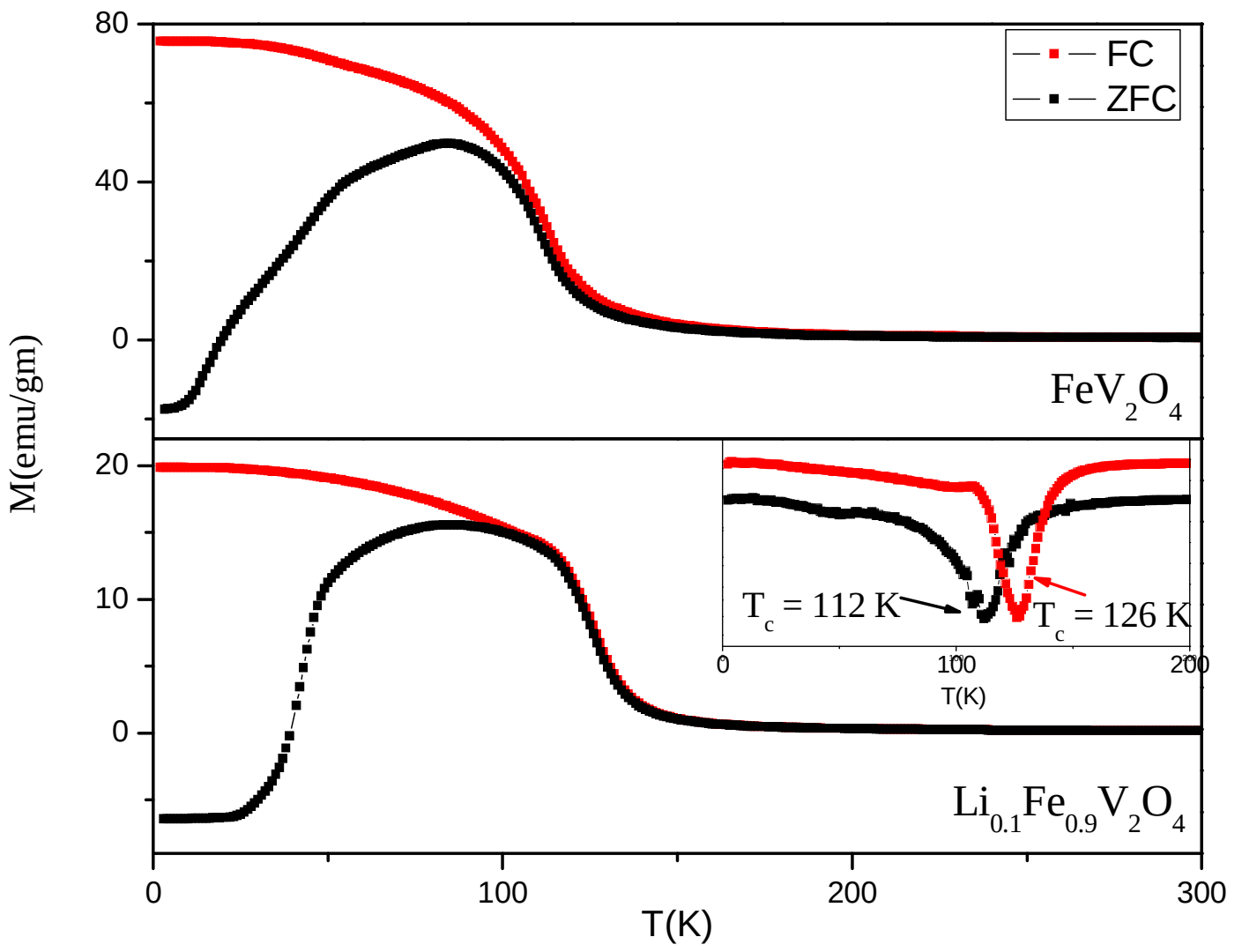

Figure 3

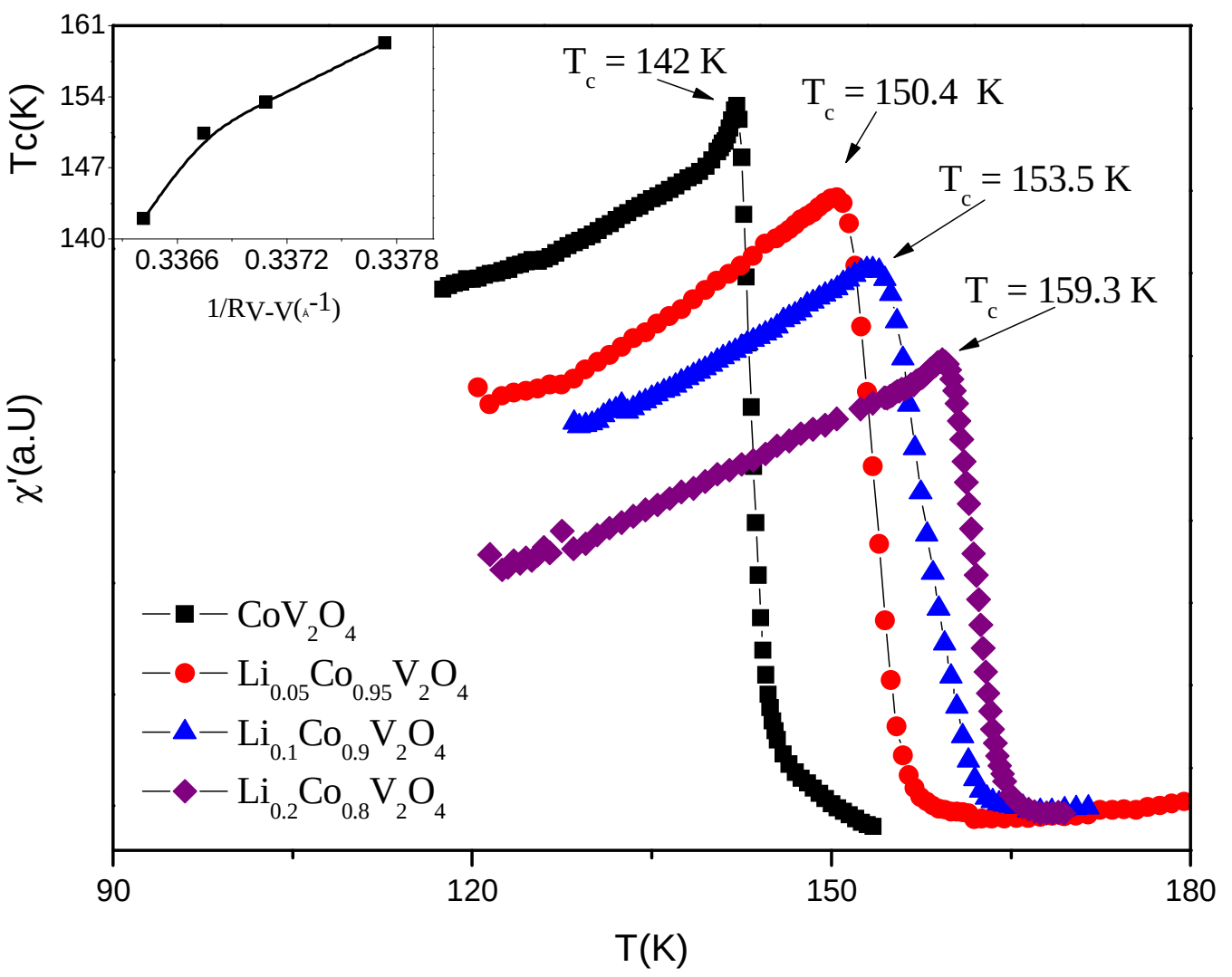

Figure 4

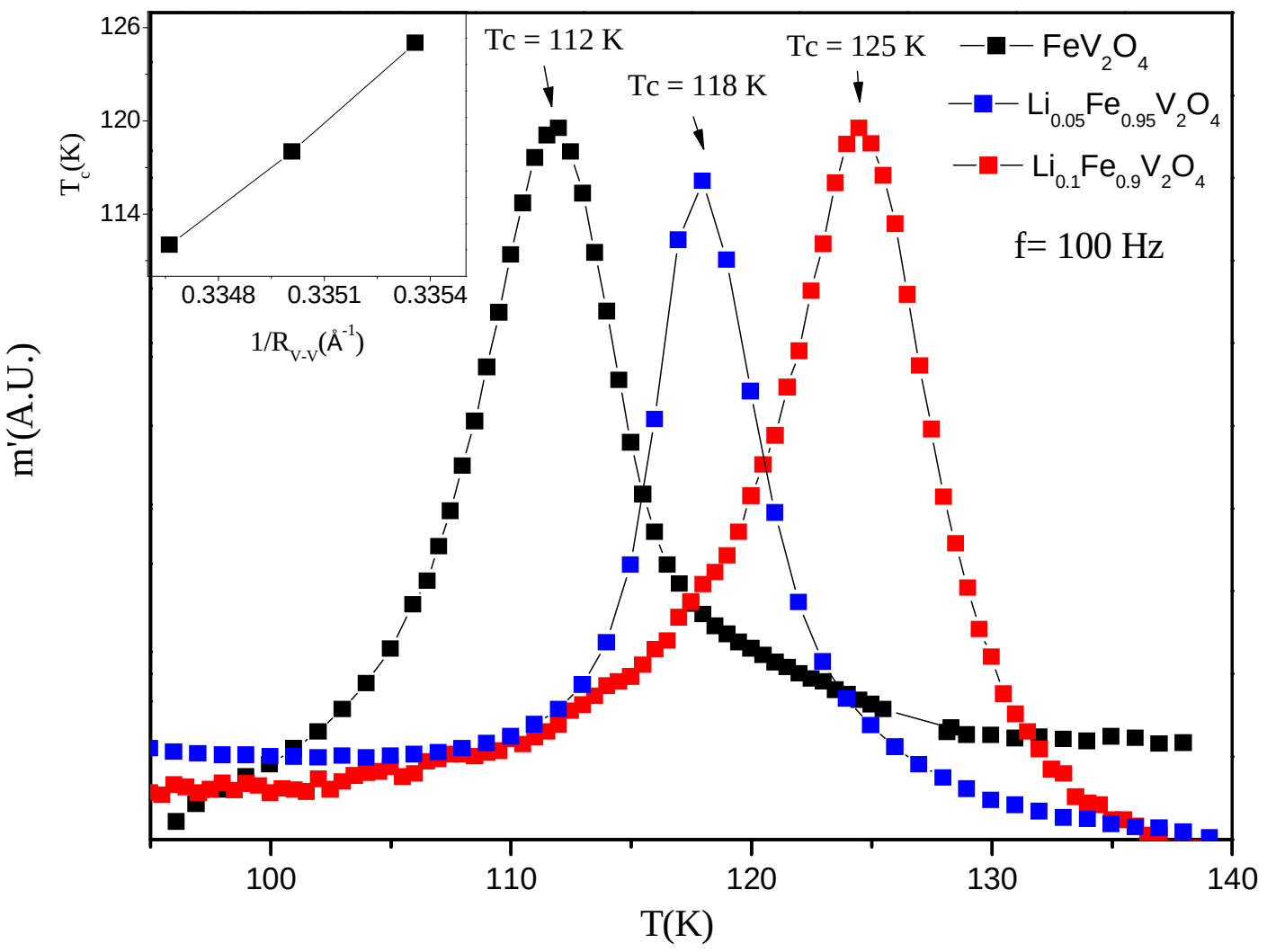

Figure 5

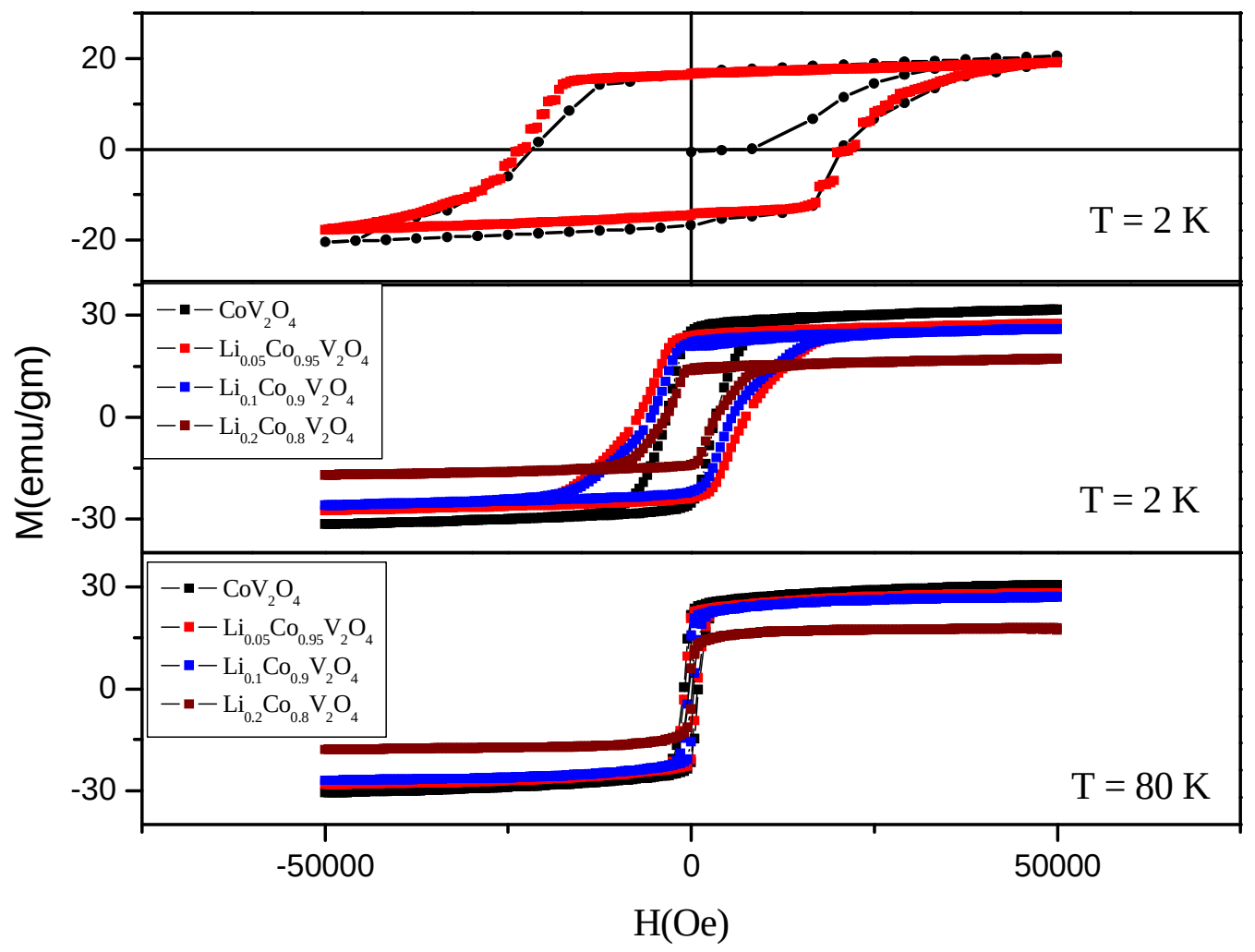

Figure 6

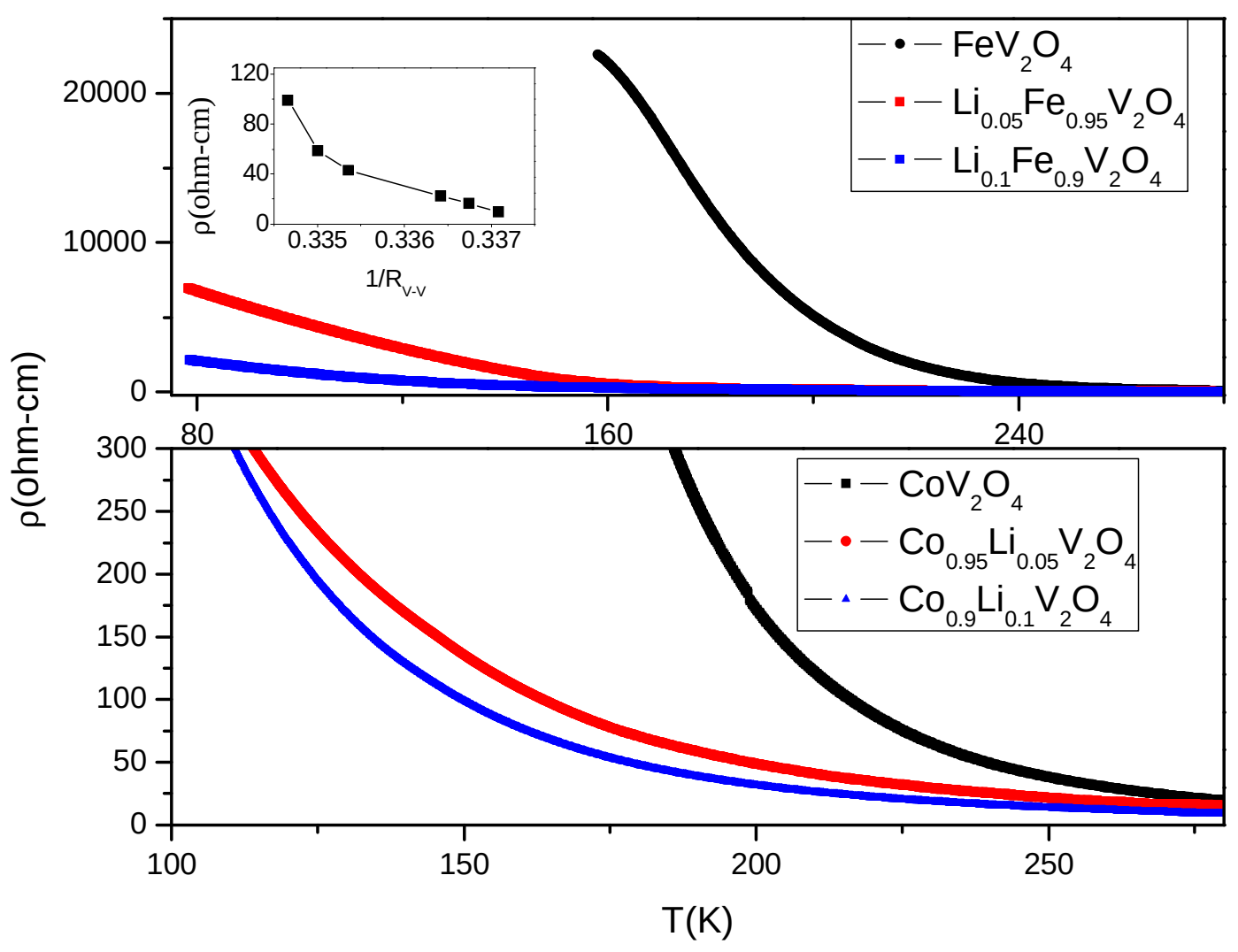

Figure 7

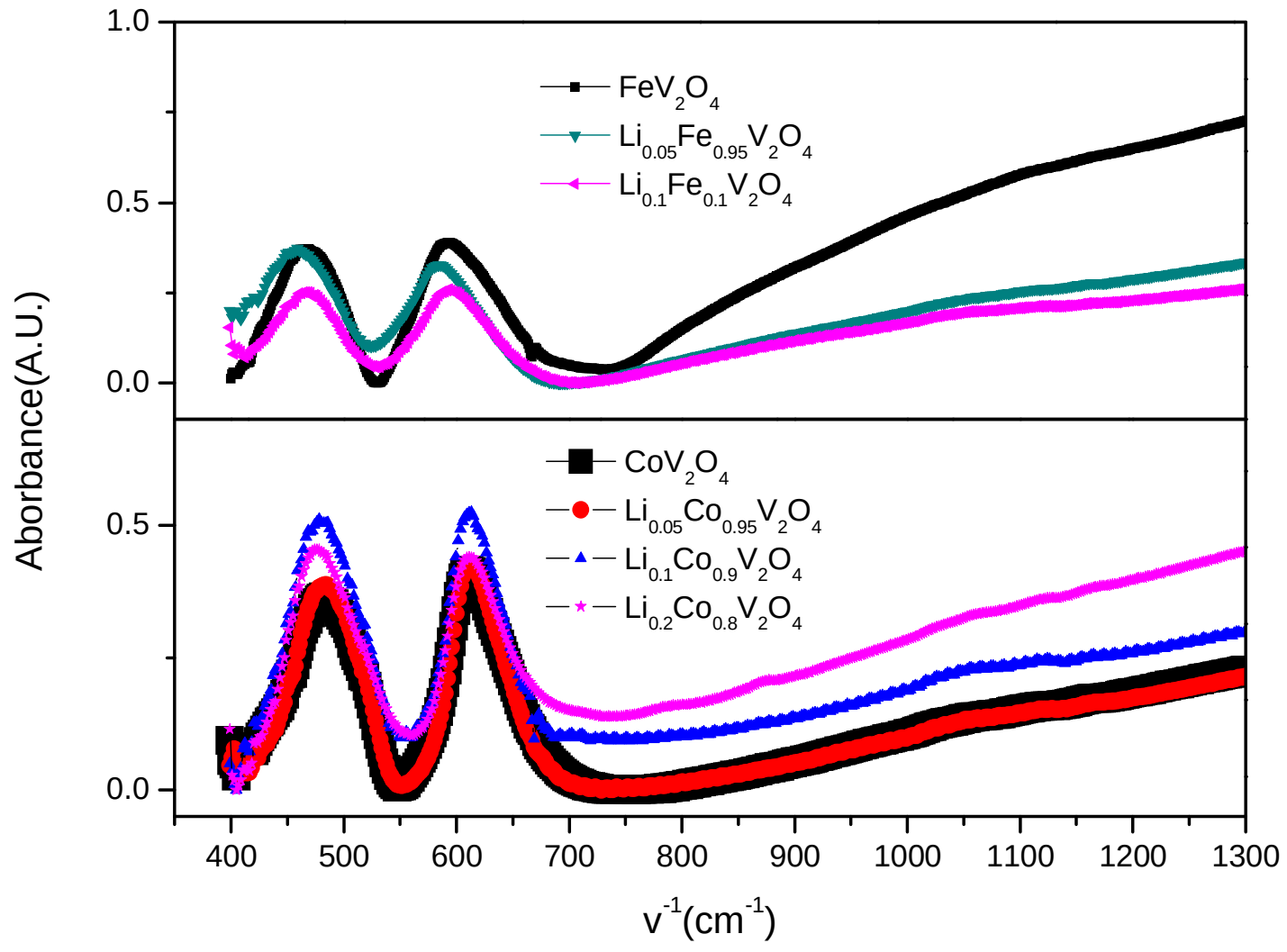

Figure 8

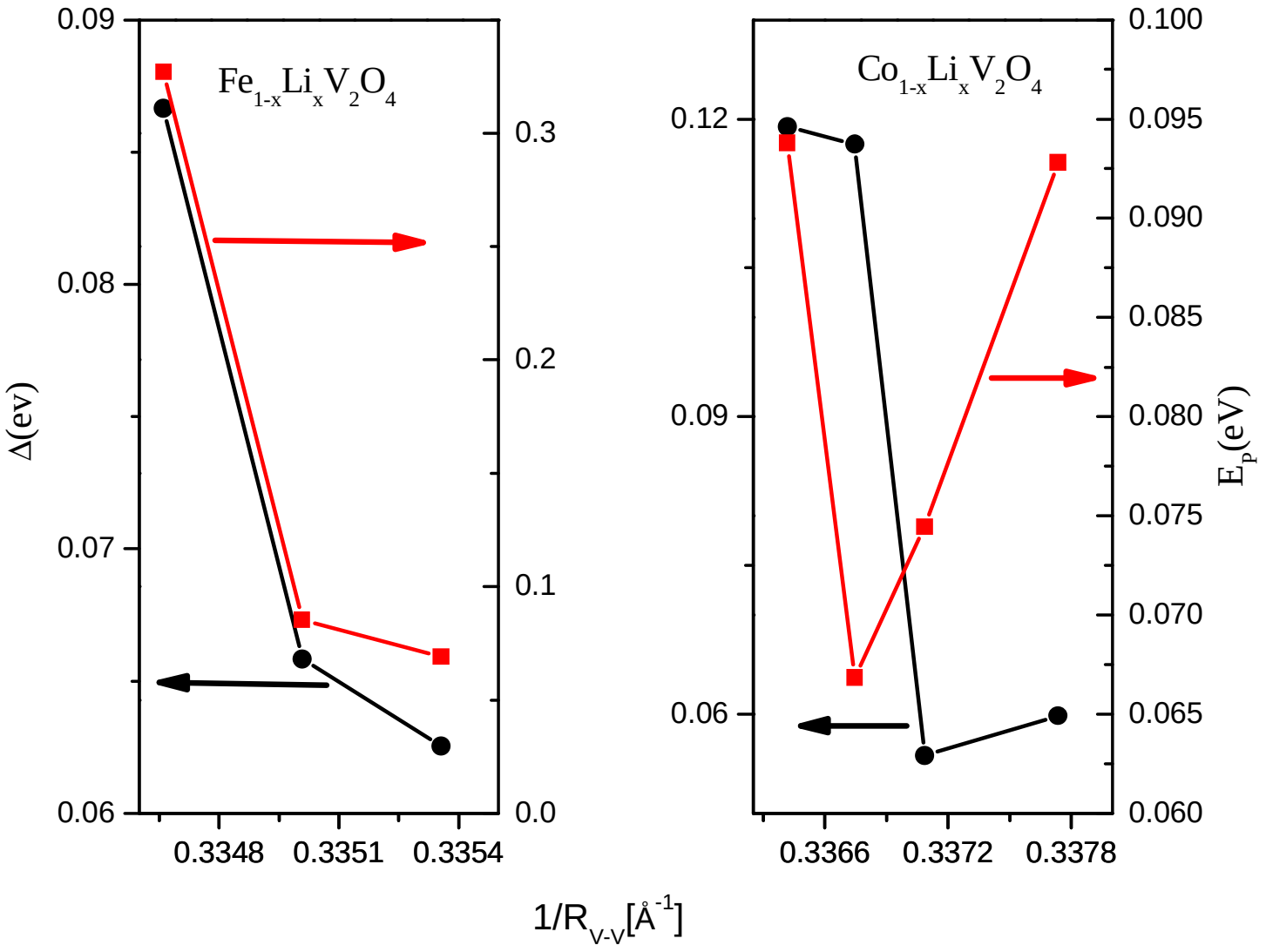

Figure 9